# Space-filling efficiency and optical properties of hemoglycin

Julie E M McGeoch[1] and Malcolm W McGeoch[2]

[1]High Energy Physics DIV, Smithsonian Astrophysical Observatory, Center for Astrophysics, Harvard & Smithsonian, 60 Garden St, MS 70, Cambridge MA 02138. USA.
[2]PLEX Corporation, 275 Martine Str, Suite 100, Fall River, MA 02723, USA

**The empty, extensive low-density lattice topology of hemoglycin is examined to understand how in space, and possibly as early as 800M years into cosmic time a rod-like polymer of glycine and iron came into dominance. A central question to be answered is whether the hemoglycin rod lattice with diamond 2H symmetry represents the most efficient covering of space by a regular arrangement of identical rods. Starting from the tetrahedral symmetry of every hemoglycin lattice vertex we find that the regular truncated tetrahedron of Archimedes may be expanded until neighboring hexagon faces are coincident, at which point space filling is 23/24 or 95.8333% complete. We describe the unit cells of the diamond 2H rod lattice and its conforming near-complete space-filling structure, which has identical symmetry. Maximum space filling via a minimum of molecular material can allow hemoglycin to drive accretion in molecular clouds, contributing to the composition of dust, and providing a background for its widespread presence in meteoritic samples and in cometary material that falls to Earth. The optical properties of hemoglycin lattice entities are derived from quantum calculations of ultraviolet and visible transition energies and strengths. The hemoglycin extinction curve duplicates the nominal 218nm ultraviolet absorption feature known as the UV bump, together with two visible absorption features present in a generic compilation of astronomical extinction data.**

**Introduction**
Hemoglycin, a specific polymer of amino acids with core mass 1494Da is found via mass spectrometry to dominate the molecular content in extracts from Allende, Acfer 086 and Orgueil meteorites [McGeoch & McGeoch 2015; McGeoch, Dikler & McGeoch 2021a; McGeoch et al. 2024a). It also is found as in-fall material in fossil stromatolites (McGeoch et al. 2024a) and cosmic dust (McGeoch & McGeoch 2024b). Its molecular absorptions, from the ultraviolet to the infrared, have been calculated and shown to provide possible explanations for well-known absorption and emission bands in the observation of stars, dust clouds and galaxies (McGeoch & McGeoch 2024c). The 220nm ultraviolet (UV) absorption "bump", well known from local measurements, has been observed in an early galaxy at z = 6.71, approximately 800My into cosmic time (Witstok et al. 2023, McGeoch & McGeoch 2024c). In hemoglycin this absorption is due to Fe(II) transitions modified by the iron-glycine bond interaction (McGeoch & McGeoch 2022; McGeoch & McGeoch 2024c). The 6.2µm emission pervasive throughout our galaxy corresponds to the Amide I transition of the anti-parallel polyglycine strands within hemoglycin (McGeoch & McGeoch 2024c). Hitherto, the UV bump and 6.2µm features have been linked with a common molecule (Blasberger at al., 2017), without a definite assignment other than that of possible polyaromatic hydrocarbons (PAHs). The slow, energetically allowed polymerization of glycine (and other amino acids) that can occur in warm, dense molecular clouds (McGeoch & McGeoch 2014) may be a first step in the formation of larger molecular groupings of polymer amide. Why one particular size and type of molecule should become so dominant throughout space and time is the question addressed.

The uniformity of the astronomical dust extinction curve with its associated UV bump has been so notable that a single generic extinction law has been able to be developed, spanning from 0.1 - 30μm [Gordon et al. 2023]. The curve was observationally featureless until a report on rocket borne stellar spectra in the vacuum ultraviolet revealed a pronounced peak in extinction at about 217nm (Stecher 1965, 1969). This feature was so well-fitted by the conduction band absorption of graphite (Stecher and Donn, 1965) that, in the absence of known alternative space-borne substances absorbing at this wavelength, graphite came to be considered an essential component of dust grains (Mathis et al. 1977). However, extinction to wavelengths shorter than 217nm was not well fitted by graphite, and it became difficult to explain the strength of extinction in terms of the projected abundance of carbon. The logic of dust model development from 1930 onward is described by (Jones et al. 2017) in a preface to the THEMIS dust modeling framework. Carbonaceous grains in the latter model are from a family of hydrogenated amorphous carbon, which is in solid state macroscopically structured networks, however nitrogen is not included whereas in (Jones, 1988) the possibility is raised of a category of polymer "organophiles" in a discussion of scattering and accretion by highly porous dust – features that together would have suggested hemoglycin. Here, we employ hemoglycin molecular analysis to construct its extinction curve ranging from the far ultraviolet through to the near infrared, finding a new origin for the UV bump in the interaction of iron with glycine.

**The existence of a regular latttice**
Hemoglycin consists of 5nm rods of glycine with iron atoms at each end. X-ray diffraction, dominated by the scattering of iron atoms, gives the spacing of the iron atoms at each end of the glycine rods (McGeoch et al. 2021b, 2022, 2024a]. The glycine part is a pair of anti-parallel, 11-glycine chains (Figure 1). This forms hemoglycin core units of mass 1494Da and 1638Da (McGeoch et al. 2021a), the latter containing an additional FeO at each end of the molecule.

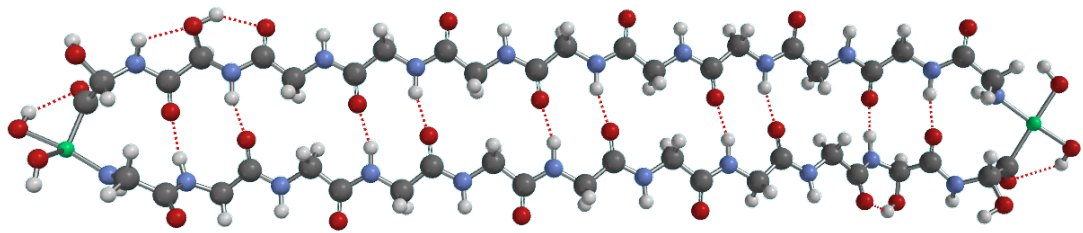

**FIGURE 1. Core 1494Da Hemoglycin, ball and stick model. Atom labels: Hydrogen white, Carbon black, Nitrogen blue, Oxygen red, Iron green.**

In meteoritic extracts (McGeoch et al. 2021a,b) including Kaba, and fossil stromatolite, there develops a structured, yet insubstantial, floating layer at the solvent interphase from which the highest concentration of the 1494Da entity is recovered (Figure 2). This layer has been interpreted as a three-dimensional lattice (McGeoch et al. 2021b) which forms from the dominant meteoritic polymer and is only seen because it floats at the intermediate density between a lower chloroform and upper water/methanol layer in the extraction vial.

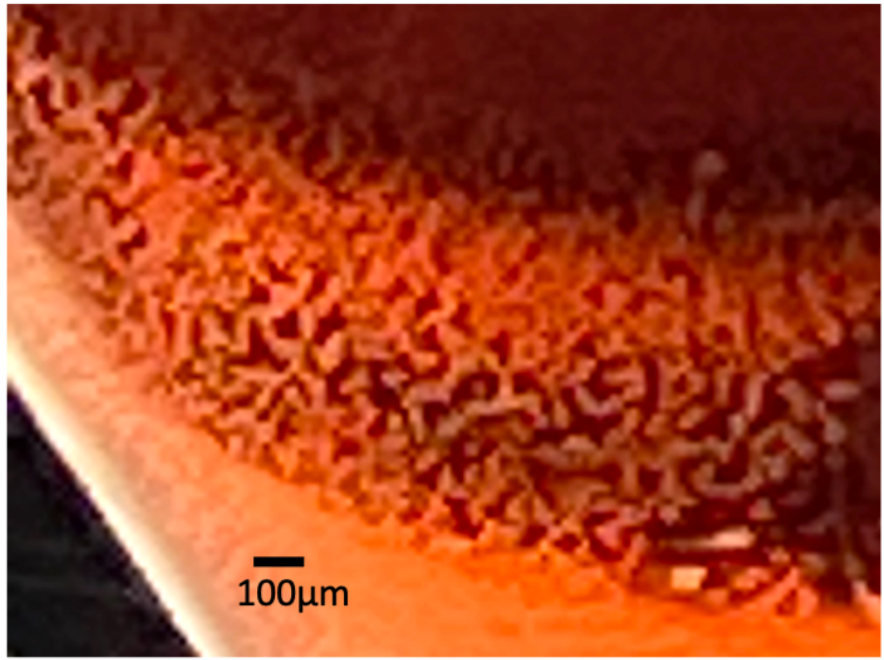

**FIGURE 2.** (reproduced from McGeoch et al. 2021b). Three dimensional macrostructures in the interphase layer of an extract from the Kaba meteorite. The red coloration is from Sudan III dye added to better visualize the structures. Similar structures are seen in Acfer 086.

The universality of hemoglycin implies a process of selection for this molecular type. The idea was proposed (McGeoch et al. 2021b) that there could be selection of the most efficient molecular type to fill space in terms of the volume of space occupied by a lattice built from a given quota of molecules. The most efficient lattice would grow ahead of less efficient lattice types because, being larger, it intercepted the largest portion of its raw materials, in this case glycine, iron and silicon. This conjecture led to our proposal of the diamond 2H structure (applied to the lattice vertices) and a preliminary discussion of its excellent space-filling efficiency, in terms of volume filled per unit of lattice rods, was presented (McGeoch et al. 2021b).

Exploratory X ray diffraction work revealed evidence of the expected tetrahedral lattice angles (Figure 3), but low-angle limitations at the 1Angstrom wavelength prevented complete structural confirmation. The modeled equilibrium molecular structure (McGeoch et al. 2022) was portrayed in space-filling orbitals (Figure 3) that indicated its very low density. The diamond 2H lattice structure of hemoglycin was later confirmed in the small calcium carbonate nodules, named ooids, that comprise most of the material in present day stromatolites (McGeoch et al. 2024a). Many chemical and physical factors were found to be in common when extracts from Orgueil and 2.3Gya fossil stromatolite were studied, leading to the expectation that the protein-like lattice already known to be within present day ooids could potentially be an extant example of the hemoglycin lattice. This turned out to be the case when a unique set of very high order diffraction rings was produced using 2Angstrom synchrotron radiation on an intact ooid (McGeoch et al. 2024a). The concentrations of iron at the lattice vertices dominated the diffraction revealing the same 4.9nm inter-vertex length already measured in crystals of Acfer 086 (McGeoch et al. 2021b) and Sutter's

Mill (McGeoch et al. 2022), together with all other inter-vertex spacings of the diamond 2H structure.

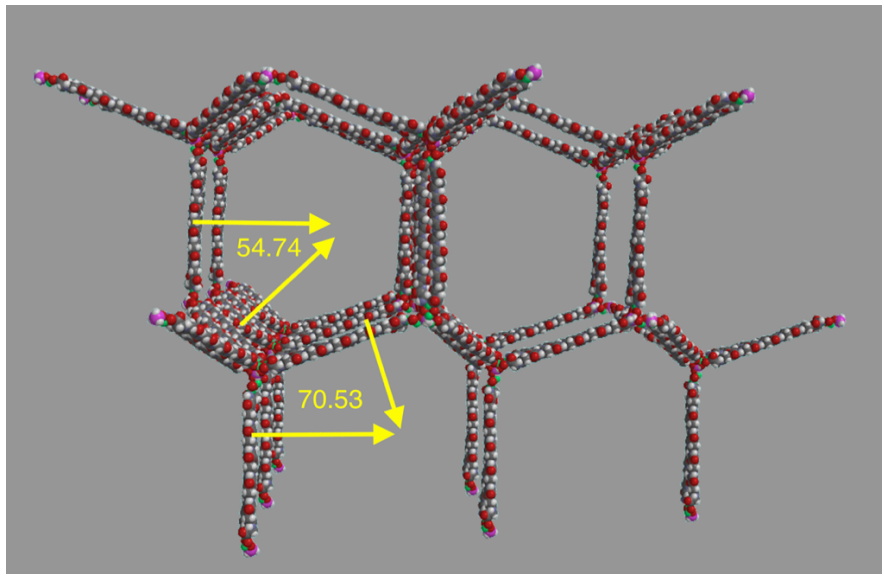

**FIGURE 3. Extensive low-density Hemoglycin lattice constructed in Spartan software (reproduced from McGeoch et al. 2022). Atom labels: Hydrogen white, Carbon black, Nitrogen blue, Oxygen red, Iron green, Silicon pink. Space-filling orbitals indicate actual scale of rods.**

**Review of Lattice data**
We briefly summarize in Table 1 the findings that led to identification of the space-filling lattice of hemoglycin. By 2017 a broad range of both biological and extra-terrestrial amino acid residues had been found in carbonaceous meteorites (Burton et al. 2012; Koga & Naraoka 2017) but no polymers apart from di-glycine (Shimoyama & Ogasawara 2002). In 2015, Folch extraction (Folch, Lees & Sloane Stanley 1957) previously used to isolate hydrophobic proteins, together with the milling of meteoritic material into micron scale particles, allowed room temperature isolation of amino acid polymers at the interphase between a lower chloroform layer and an upper water-methanol layer (McGeoch et al. 2015). MALDI mass spectrometry avoided the necessity for chemical pre-treatment and chromatography, giving access to the unmodified polymer amide content of meteorites. A volcanic rock sample collected from a lava flow in Hawaii provided a control without organic molecules. The polymers first identified ranged up to 20 amino acid residues, dominated by glycine with about 20% of the glycine residues oxidized to hydroxyglycine. Further experiments in samples from the Acfer 086 and Allende meteorites (McGeoch & McGeoch 2017) produced a surprising 4,641Da molecular peak accompanied by multiples up to 18.6 kDa. This peak was later identified as a trimer of the 1494Da "core" polymer unit (McGeoch et al. 2021a) with three core units joined by Si atoms at the vertex of a triskelion. This indicated that a planar hexagonal lattice could be one of the lattice forms of this core unit, now called hemoglycin on account its iron content.

X-ray diffraction of fiber-like crystals from extracts of Acfer 086 revealed a characteristic length of 5nm for a planar lattice that was curved, or rolled up, within the fibers (McGeoch et al. 2021b).

Physical forms floating at the Folch interphase (Figure 2) of this and other meteoritic extracts suggested the existence of a three-dimensional molecular lattice composed of hemoglycin rods (McGeoch et al. 2021b). The 120$^O$ angles observed in the forms were consistent with the orientation of clear passages in a trigonal lattice of the diamond 2H type (Raffy, Furthmuller & Bechstedt 2002) and analysis was performed (McGeoch et al. 2021b) that showed highly efficient space-filling (i.e. volume filled by the lattice per rod of polymer) for this symmetry. In 2024 hemoglycin was found to be present in fossil and present-day stromatolites (McGeoch et al. 2024a) and in present day stromatolite ooids longer wavelength (2Angstrom) X-ray scattering from the iron vertex clusters yielded many higher order diffraction rings that confirmed a diamond 2H structure for the lattice vertices (McGeoch et al. 2024a). The rods in this case were 1638Da versions (McGeoch et al. 2021a) of the 1494Da unit, with additional iron and oxygen atoms as shown in Figure 4. With Si at the vertices, there is a guiding tendency toward tetrahedral symmetry. Additional geometrical views of this lattice are given in Figure 5, showing the symmetry features further described in the following sections.

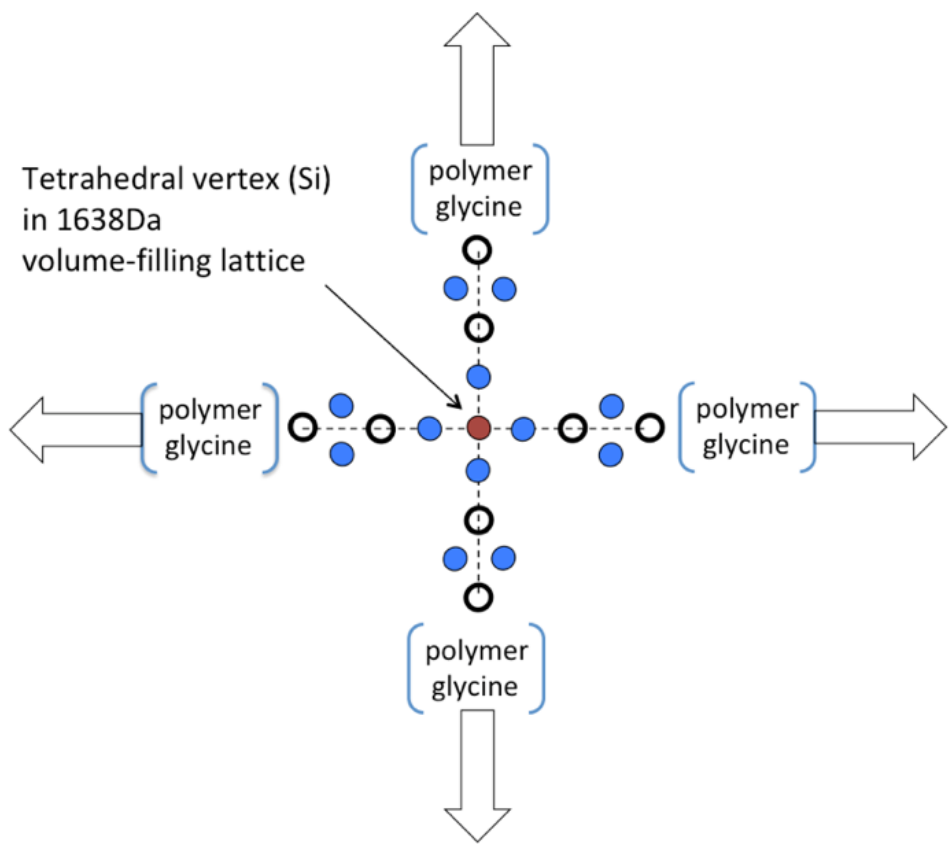

**FIGURE 4. 1638 Da vertex with tetrahedral symmetry. Si red, O blue and Fe open circles, with Fe bonded to the antiparallel glycine chains.**

Lastly in Table 1, evidence for the 1638Da version of hemoglycin (that still has 22 glycine residues) is seen in sea foam (McGeoch et al. 2024b), where it could be contributing to a planar lattice that imparts strength to cell walls in the foam. In this case the source is believed to be cosmic dust which continually falls to Earth from space (Rojas et al. 2021). In addition to the synopsis of lattice results in Table 1, summaries of the isotope and hemoglycin light emission and absorption work are given in the Supplementary Data as Table S1 and Table S2, respectively.

**Table 1. Data supporting Hemoglycin being a low-density Lattice**

| DATE PUBLISHED | SOURCE | EXPERIMENT | RESULT |
|---|---|---|---|
| 2015 https://onlinelibrary.wiley.com/doi/10.1111/maps.12558 | Allende Murchison meteorites Hawaii volcano control | 1. MALDI/TOF mass spectrometry 2. Amino acid analysis of Folch extract | Glycine and hydroxyglycine polymer present in Allende, up to 20 residues |
| 2017 https://arxiv.org/pdf/1707.09080.pdf | Acfer-086 Allende Hawaii volcano control | MALDI/TOF mass spectrometry | 4,641Da molecule with multiples up to 18,056Da |
| 2021 https://arxiv.org/abs/2102.10700. [physics.chem-ph] | Acfer-086 Allende Kaba | MALDI/TOF mass spectrometry | Hemoglycin of mass 1494Da is of glycine, hydroxy-glycine, Fe and O. Hemoglycin is connected covalently in triplets by silicon to form a 4,641Da triskelion. |
| 2021 https://aip.scitation.org/doi/10.1063/5.0054860. | Acfer-086 | X-ray diffraction APS synchrotron | Fiber pattern gave polymer length of 5nm. Low density structures at Folch interphase. **Space filling 3D lattice discovered.** |
| 2024 https://doi.org/10.1017/S1473550424000168 | Fossil and present-day stromatolites | 1. MALDI/TOF MS 2. X-ray diffraction APS and Diamond Light Source synchrotrons 3. FTIR amide band. | Stromatolites contain the same molecule, hemoglycin, as carbonaceous meteorites. **Diamond 2H 3D lattice confirmed in X-ray diffraction.** |
| 2024 https://doi.org/10.1039/d4ra06881e | Sea Foam | MALDI /TOF mass spectrometry | Sea Foam contains the same molecule, hemoglycin, as meteorites and stromatolites |

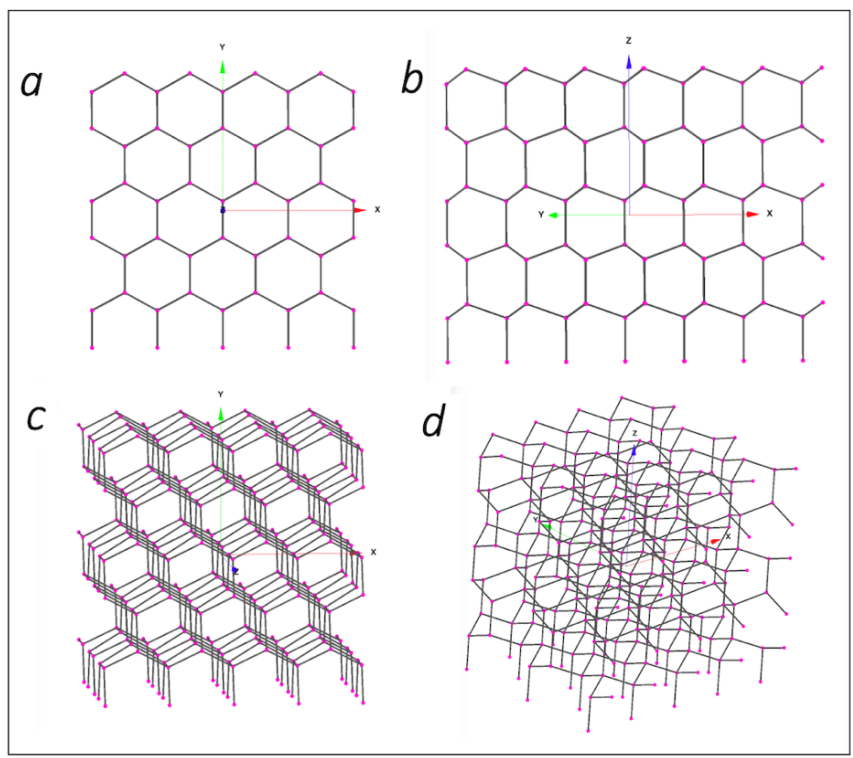

**FIGURE 5. Views of the hemoglycin space-filling lattice. Each connecting rod is a 5nm polymer comprising anti-parallel glycine chains terminated by a grouping of oxygen and iron atoms. The vertices, shown at enlarged scale, carry silicon atoms (pink). (*a*) View directly down the trigonal symmetry axis; (*b*) one of six clear views perpendicular to the symmetry axis; (*c*) similar to (*a*), except slightly offset; (*d*) angle view showing puckered hexagons at the top.**

**Hemoglycin as an efficient space-filling lattice.**
Here we study the nature of space-filling by the diamond 2H lattice to establish whether this is the most efficient possible lattice for space filling by "regular" connecting rods, that is rods of a single chemical structure and length, with inversion symmetry around a central point, giving end-for-end symmetry to ensure maximum simplicity during lattice assembly. Considering a lattice with $N$ rods converging at each vertex, an intuitive measure of space-filling comes from the construction of a plane perpendicular to each rod that intersects each rod at the same distance away from the vertex. If these planes are extended away from a rod until they just intersect each other, the enclosed volume can be compared to the number of rods required for its creation. For $N = 3$ it is not possible to enclose any volume. For $N = 4$ the planes enclose a tetrahedron and for $N = 6$ the planes can enclose a cube. In (McGeoch et al. 2021b) we show that the lattice volume enclosed per rod at $N = 6$ is 2.3 times less than at $N = 4$, which might indicate more than sufficient rods converging at each vertex. We look further into tetrahedral space filling to understand the factors that contribute to its high efficiency.

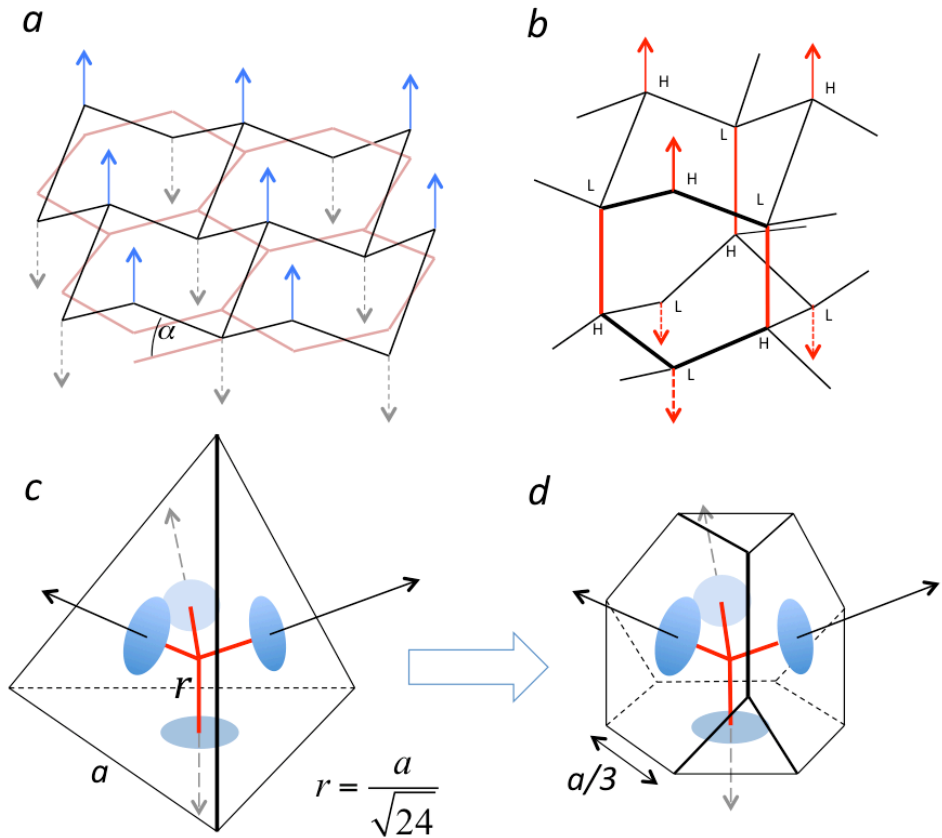

**FIGURE 6. Construction of the diamond 2H lattice and volume-defining surfaces.** *a* **Plane containing flat hexagons (beige) with rotated edges (black).** *b* **Stack of two "puckered" hexagonal forms with connecting vertical rods.** *c* **constructed tetrahedron with faces at right angles to lattice rods.** *d***. Truncation of the tetrahedron with regular edge length.**

The diamond 2H lattice can be constructed simply as follows: With reference to Figure 6(*a*) each edge of the plane hexagons is rotated by an angle $\alpha$ around an axis in the hexagon plane lying perpendicular to the edge at its center. This has the effect of "puckering" the hexagon to create three "high" points and three "low" points (black lines in Fig 6*a*). Figure 6b shows two puckered hexagons, one above, one below, linked by vertical rods ascending from "high" points on the lower form to the corresponding "low" point in the form above, which has rotated $60^O$ about a vertical axis relative to the identical form below. All of the rod connections are of the same length *h*. In (McGeoch et al. 2021b) it is shown that the volume defined by this lattice is maximized for $\sin\alpha = 1/3$, or $\alpha = 19.471$ deg. which defines exact tetrahedral symmetry at every vertex. This is an optimum for the space filling of this construct, but further insight comes from the tetrahedral constructions in Fig. 6c and 6d. The vertex lies at the center of the "in-sphere" of the tetrahedron, which touches the center of each face and has radius $r = a/\sqrt{24}$ where *a* is the edge length of the tetrahedron.

When we increase *r*, our full tetrahedron collides with similar volumes expanding from neighboring vertices, suggesting use of a modified, truncated, form. The regular truncated tetrahedron of Archimedes (TTA) (Fig. 6d), with each edge equal to $a/3$ is quickly seen to yield improved space filling without interferences. When $r = h/2$ the hexagonal TTA faces around one vertex sit exactly onto the TTA faces around all neighboring vertices leaving very little space still to be filled.

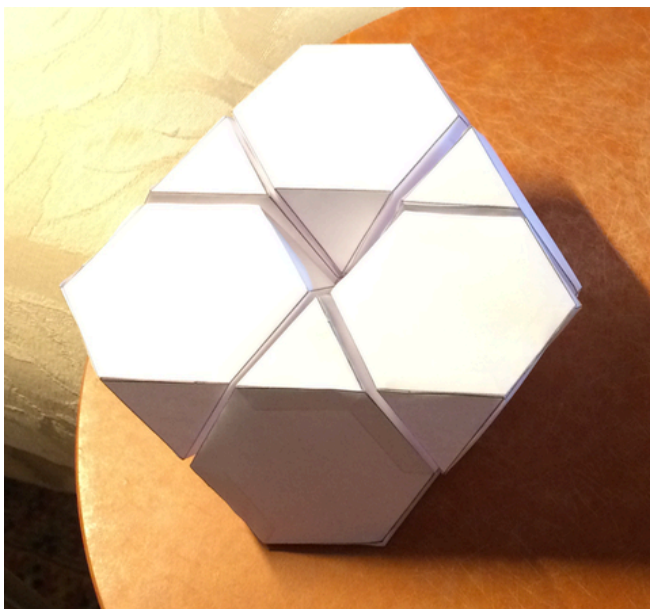

**FIGURE 7. Six TTAs positioned with the centers of their in-spheres at the six tetrahedral vertices of one lattice hexagon.**

The packing corresponding to the six vertices of one puckered hexagon is illustrated in Figure 7 with paper TTA models. Alternate TTAs point up and down, following the "low" and "high" alternation of the vertices in the lattice hexagon. For example, the TTA at lower front center of the stack represents a "low" vertex, while those to the left and right of it represent "high" vertices. There is only a small un-filled region at top center, which is a regular tetrahedron with edge $a/3$. This only goes half-way down toward the base. Continuing downward we enter a similar tetrahedral opening to the bottom of the stack (or "layer") with the difference that the lower tetrahedron is rotated $60^O$ relative to the upper one. As expected from the construction of Fig. 6a and 6b this packing extends horizontally without break, the only exception being the unfilled $a/3$ tetrahedrons. An extended layer is shown in Figure 8.

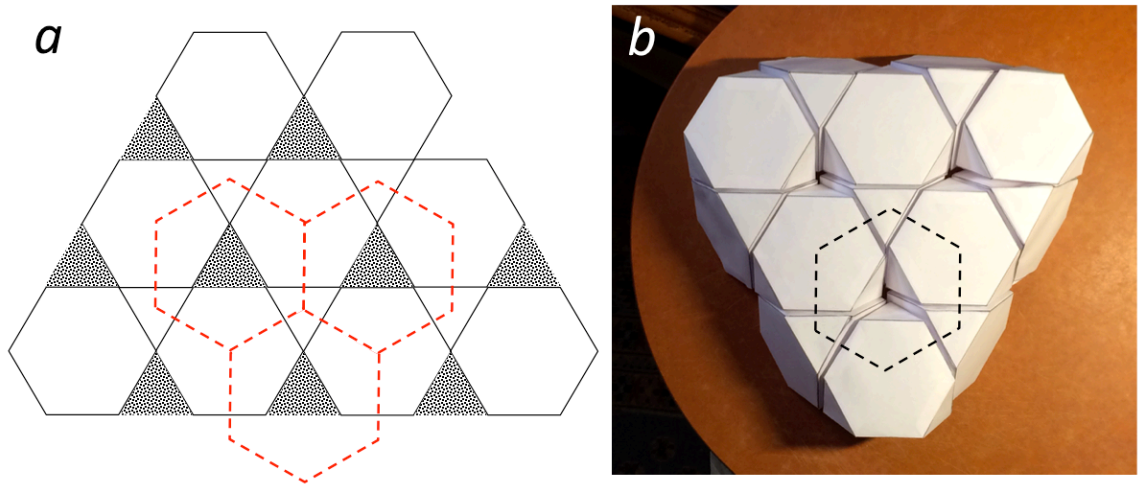

**FIGURE 8.** Extended layer showing repeats. Grey triangles are empty small tetrahedrons. Dashed lines locate the position of lattice hexagons viewed from above.

**Table 2.** Summary of geometrical relationships for the most efficient TTA packing with $r = h/2$, when constructed TTA's touch at the mid-position of diamond 2H lattice rods.

| Parameter | "h units" | "a units" |
|---|---|---|
| Rod length | $h$ | $a/\sqrt{6}$ |
| Packing layer thickness $L$ | $4h/3$ | $2\sqrt{2}a/3\sqrt{3}$ |
| Area of lattice hexagon $A_H$ | $4\sqrt{3}h^2/3$ | $2\sqrt{3}a^2/9$ |
| Depth of a packing unit cell $2L$ | $8h/3$ | $4\sqrt{2}a/3\sqrt{3}$ |
| Volume of a lattice half-cell $LA_H$ | $16\sqrt{3}h^3/9$ | $8a^3/27\sqrt{2}$ |
| Volume of tetrahedron of edge $a$   $V_T(a)$ | $\sqrt{3}h^3$ | $a^3/6\sqrt{2}$ |
| Volume of tetrahedron of edge $a/3$   $V_T(a/3)$ |  | $a^3/162\sqrt{2}$ |
| Fraction of space filled by constructed TTAs on diamond 2H rod lattice $\left[LA_H - 2V_T(a/3)\right]/LA_H$ | $\dfrac{23}{24} = 95.8333\%$ | |

The packing efficiency of one layer as shown is 23/24 = 0.958333, according to formulae in Table 2. The vertical pattern repeat, however, involves two layers in accordance with the rotation between puckered rod hexagons in Figure 6b. A third hexagon if added in Fig. 6b is identical to the lower, first hexagon. In 6b, the upper hexagon can be created via a mirror image of the lower hexagon in a mirror lying horizontally through the mid-points of the vertical rods joining the hexagons. Likewise, a mirror image of the space-filling TTA form in Figure 7 via a mirror lying in its upper surface, will complete a space-filling "unit cell". Further stacking of unit cells fills all of space with the exception of 1/24 of its volume, as may be seen from the geometrical relationships summarized in Table 2. The lattice half-cell referred to in Table 2 corresponds to the "quasi-cell" used in (McGeoch et al. 2021b) to simplify the discussion. The same packing efficiency of 23/24 for TTAs was previously derived (Jiao & Torquato 2011) using a Bravais lattice development of inversion-symmetric pairs of TTAs in a different approach from the present starting point of the rod-like lattice of diamond 2H symmetry.

**Hemoglycin as a component of cosmic dust**

In the present section we first test whether a large dust component of hemoglycin would be consistent with elemental abundances. Secondly, the optical properties of hemoglycin dust particles are derived and compared with astronomical extinction data.

**Consistency with metallic abundances**

If, as we have supposed, the growth (in a molecular cloud) of the hemoglycin lattice has outstripped that of other lattice designs on account of its space-filling efficiency, then a corollary is that hemoglycin lattices will be a dominant "dust" component. This proposition is not in conflict with available data if the open lattice contains a partial fill of the array of small-molecule ices proposed to fit the dust absorption and scattering spectrum of molecular clouds (Dartois et al, 2024). One question that is easily answered relates to the elemental composition of hemoglycin, whether this matches the solar system abundances, which ought to be the case when it dominates the core accretion within our protoplanetary disc (Aikawa & Herbst 1999). In Table 3 the elements of the 1638Da lattice rod ($C_{44}O_{32}N_{22}H_{66}Fe_4$) plus the effective half Si atom that each rod carries in relation to its two vertices, are ratio-ed to the solar abundances of these elements (Anders & Grevesse 1989). For example, $C_H/C_S$ is the ratio of carbon in hemoglycin to carbon in the solar system, with a scaling factor applied that sets $N_H/N_S = 1.00$ .

**Table 3. Scaled ratios of elements in hemoglycin to solar abundances.**

| Ratio | $C_H/C_S$ | $N_H/N_S$ | $O_H/O_S$ | $Fe_H/Fe_S$ | $Si_H/Si_S$ |
|---|---|---|---|---|---|
| Scaled ratio | 0.62 | 1.00 | 0.19 | 0.63 | 0.07 |

One way to read this is to suppose for the sake of argument that 100% of solar system nitrogen is bound up in hemoglycin. Then 62% of solar system carbon, 19% of O, 63% of Fe and 7% of Si would also be bound up in hemoglycin. There is thus a rough consistency between the elemental composition of hemoglycin and that of the solar system, the main solar system excesses relating to O and Si. Certainly, this 100% scenario is not the case, but it illustrates the possibility of potentially a large fraction of the dust in at least our own proto-stellar disc being composed of hemoglycin.

**Optical properties and extinction**

Although we have measurements for hemoglycin of visible [McGeoch et al. 2022] and infrared [McGeoch et al. 2024a] absorptions in meteoritic extract crystals and stromatolite ooids, we do not have data on absorption in a gas of the free molecule or the extended empty lattice in the laboratory. Besides, our visible region absorption measurement only extends down to 300nm, whereas the most interesting dust absorption feature is the "UV bump" in the extinction curve at nominal 218nm. Here we approximate the lattice behavior by considering an array of core hemoglycin molecules (at 1494Da) disposed at the angles of the rods in the diamond 2H lattice. Using the Spartan program [Wavefunction Inc. 2024] that embodies the Q-Chem [Epifanovsky et al. 2021] quantum chemistry solver we can explore via theory the whole range of absorptions from 0.1μm to 30μm. This program predicts bound-bound absorptions between stable molecular states, specifically upward transitions from the ground electronic state. In an experimental study [Iwanami & Oda 1983] of the vacuum ultraviolet absorption of polymer amino acids, evidence has been seen for a large set of neutral molecular states above the ionization potential, apparently associated with the promotion of an electron into "conduction" states within the polymer backbone, an effect not included in the bound-bound calculations made here, providing a limit to accuracy for wavelengths below about 160nm, although in practice the program has predicted bound-bound absorptions down to 145nm.

Dust in molecular clouds within the interstellar medium is primarily observed via its modification of stellar spectra. Its large and smoothly increasing ultraviolet attenuation [see Jones 2017, Gordon 2023, Wikipedia 2025 for an entry to the extensive literature on dust] is consistent with the presence of small particles that need not necessarily absorb light but scatter according to the well-known relation [Jackson 2001, eqn. 10.11] for spherical particles:

$$\sigma = \frac{8\pi}{3} k^4 a^6 \left| \frac{\varepsilon_R - 1}{\varepsilon_R + 2} \right|^2 \qquad (1)$$

Here $\sigma$ is the total scattering cross section, $k=2\pi/\lambda$ is the wavenumber associated with light of wavelength $\lambda$, $a$ is the particle radius, and $\varepsilon_R$ is the relative dielectric constant of the particle material which is assumed isotropic. Although we refer to a fragment of lattice as a particle, its shape need not be definite. In the following we calculate absorption and scattering for an assumed approximately spherical piece of free-floating lattice that we refer to as a "ball", with radius $a$. The inverse dependence on wavelength to the fourth power (via $k$) is responsible for the steep increase in scattering toward the ultraviolet, but the even stronger dependence upon $a$, the particle radius, has its origin in the approximation $a \ll \lambda$ made in the derivation. When this is the case all induced dipoles within the material (at the molecular level) oscillate in phase, and their radiated power is then proportional to the square of the number of dipoles. The number of dipoles goes as the volume of the sphere $4\pi a^3/3$, hence the resulting sixth power dependence of the scattered power on radius $a$. The strength of individual dipoles is captured in the relative dielectric constant term $\varepsilon_R$.

All materials have characteristic optical absorptions, which add further attenuation depending upon their wavelength. In the presence of absorption [Jackson 2001, eqn. 7.51] the relative dielectric constant $\varepsilon_R$ is a complex number, the real part associated with the material refractive index, and the imaginary part ($i$) with optical absorption:

$$\varepsilon_R = \frac{\varepsilon(\omega)}{\varepsilon_0} = 1 + \frac{Ne^2}{\varepsilon_0 m} \sum_j f_j \left( \omega_j^2 - \omega^2 - i\omega\gamma_j \right)^{-1} \qquad (2)$$

where the incident light angular frequency is $\omega = kc$ and the sum is over the different transitions of oscillator strength $f_j$ that $Z$ electrons within the molecule can participate in, with angular frequencies $\omega_j$ and damping coefficient $\gamma_j$ that gives rise to absorption. $N$ is the number of molecules per unit volume, $e$ the electron charge, $\varepsilon_0$ the permeability of free space and $m$ the electron mass. For hemoglycin 4.9nm rods (molecules) in the diamond 2H lattice $N = 1.104 \times 10^{25}$ rods/m$^3$.

When the dust particle is an empty hemoglycin lattice ball of "radius" *a*, there are molecular absorptions as previously discussed. The 480nm chiral absorption was predicted in quantum chemistry [McGeoch et al. 2022] and confirmed experimentally close to the expected wavelength. Similar calculations are here performed that extend further into the ultraviolet, and cover the UV -> visible -> near IR range from 140nm to 2µm. The results for transition wavelengths $\lambda_j$ and strengths $f_j$ in our principal calculation are listed in Table 4. "Spartan" quantum chemistry software was used, in Hartree-Fock 3-21G basis with Random Phase Approximation (RPA) output. RPA had previously been found to better represent absorption data in the visible, compared to Single Excitation Configuration Interaction (CIS) output. In these calculations the full 1638Da rod with 4 Fe atoms was in general unable to converge, due to the very high density of Fe states. We used the 1494Da core in its place, and therefore the silicate vertex (Figure 4) along with four of the Fe atoms was not included. Tests with this latter fragment alone produced the well-known silicate infrared absorptions in the region of 10microns.

**Table 4. Calculated RPA transition wavelengths and strengths for the 1494Da hemoglycin core with 2 R-chirality hydroxyglycine residues at C terminus on Fe at one end and 2 S-chirality hydroxyglycine residues at C terminus on Fe at other end. Numbers in bold represent the strongest visible and near-ultraviolet transitions. Numbers in italics show all transitions with strength >0.1 for a polymer backbone with no Fe, providing data below 148nm.**

| Wavelength (nm) | Oscillator Strength | Wavelength (nm) | Oscillator Strength |
|---|---|---|---|
| 1455.4 | 0.000074 | **218.64** | **0.316417** |
| 1451.8 | 0.000126 | 210.36 | 0.047928 |
| 1116.3 | 0.000194 | 200.70 | 0.022901 |
| 993.33 | 0.000197 | 200.31 | 0.051230 |
| 766.63 | 0.000176 | 197.79 | 0.075298 |
| 739.72 | 0.000073 | 197.30 | 0.094989 |
| **646.42** | **0.009703** | 193.74 | 0.003738 |
| **477.37** | **0.010517** | 192.47 | 0.001882 |
| 398.98 | 0.000226 | 186.03 | 0.000955 |
| 393.19 | 0.000317 | 183.27 | 0.003352 |
| 384.98 | 0.006970 | 151.50 | 0.106445 |
| 334.69 | 0.004521 | 150.09 | 0.031994 |
| 328.96 | 0.006293 | 148.56 | 0.020008 |
| 324.79 | 0.005032 | 148.39 | 0.295151 |
| 323.00 | 0.003207 | 147.89 | 0.288224 |
| 270.60 | 0.002594 | *146.59* | *1.928227* |
| 266.66 | 0.017277 | *145.93* | *1.066054* |
| 238.76 | 0.079598 | *145.79* | *0.358401* |
| 229.12 | 0.037800 | *145.33* | *0.388388* |
| **228.90** | **0.141131** | *144.76* | *1.205900* |

Several assumptions are made when using expression (2). Firstly, the isotropic medium assumed in the derivation of (1) has induced molecular dipoles that are randomly oriented whereas a diamond 2H hemoglycin lattice does not. We explored this in a detailed calculation of the net induced dipole radiation from one class of dipoles aligned with the lattice rods and found, surprisingly, that the scattered radiation from a unit cell of the lattice was identical in its distribution and polarization to the scattered radiation of an isotropic medium (Section S3), a result we attribute to the high symmetry of the lattice. Although we have not extended this calculation to the full dielectric tensor we will assume that this result extends to all induced dipoles for the present discussion, enabling us to use randomly averaged transition strengths directly from the RPA output. Secondly, it is assumed that the internal electric field is very little different from the incident external field, which is a good approximation in this low-density lattice. Thirdly, (2) is only valid for relatively low fractional absorption as radiation transits the dust particle, i.e. the dust particle is not at all close to being "black" to incident light at the center wavelength of any absorbing transition.

Although the quantum calculation gives oscillator strengths $f_j$ and center frequencies $\omega_j$ it cannot without much added complexity account for the vibrational and rotational transitions of the molecule, which can contribute to $\gamma_j$, the "damping constant" that gives spectral width to a transition. There is empirical damping data at 483nm [McGeoch et al. 2022], where the measured transition width is approximately one tenth of the central frequency. Approaching the ultraviolet, there is a rapidly increasing density of states that can accept an electron promoted by absorption, giving rise in principle to an increased damping rate, and in turn driving the physics toward the dominance of absorption over re-radiation (scattering) at wavelengths corresponding to a transition. Here we will introduce the very simple assumption that damping rates are linearly proportional to transition frequency, setting the full width at half maximum of a transition $\gamma_j = \alpha_D \omega_j$ where $\alpha_D$ is a constant factor of the order of 0.1. This assumption is approximately validated by the fitted damping constants to the general extinction model of [Gordon et al. 2023]. Subsequent to absorption in hemoglycin there can be visible re-radiation from high-level states [McGeoch et al. 2023] and, in another side effect, this decay in turn can populate the upper level of the hemoglycin 6.2μm vibrational transition that may be the source of the widely observed cosmic emission at that wavelength [McGeoch et al. 2024c].

Figure 9 shows calculated extinction cross sections (in units of m$^2$) for an empty hemoglycin lattice "ball" of 100nm radius, with damping constants 0.2-times or 0.1-times the transition frequencies. The cross section for other radii (keeping $a << \lambda$) is found via multiplication of the plotted data by $(a/100\text{nm})^6$.

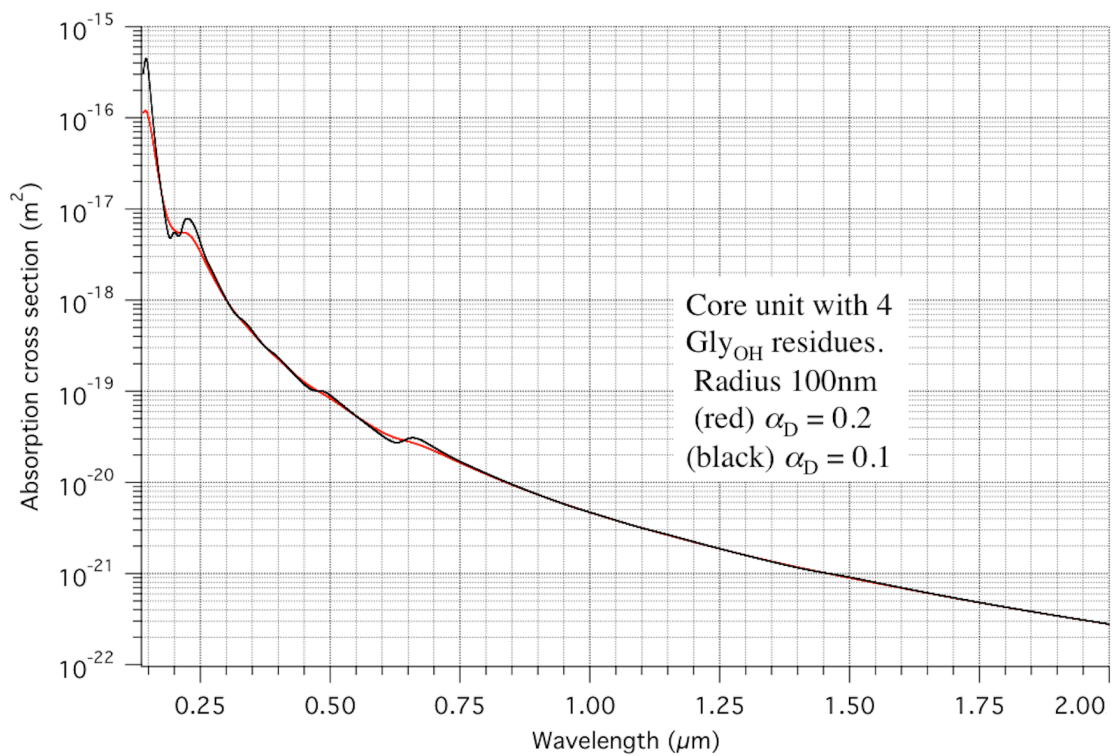

**Figure 9. Calculated extinction cross section (m$^2$) vs wavelength (µm) for an empty 100nm radius hemoglycin lattice ball, with lesser damping (black curve) and greater damping (red curve). Core unit at 1494Da with 2Gly$_{OH}$ at the C-terminus of each antiparallel strand.**

The most prominent feature on these curves is the "bump" at nominal 220nm, which is produced by the interaction of an iron atom at one or other end of the core molecule with the C-amino terminals of the attached glycine chains. We deduced in a more detailed study of this interaction [McGeoch et al. 2022] that visible and near ultraviolet molecular transitions in hemoglycin had their origin in allowed Fe(II) transitions. Fe within the molecule had a Mulliken charge of +1, and there was a perfect correlation of the calculated <u>molecular</u> transition energies with Fe(II) <u>atomic</u> transitions out of the near-ground Fe(II) $a^4F_{9/2}$ state. Of the two standard Spartan outputs (in the CIS and RPA approximations) we found that the RPA results agreed closely with the measured 483nm absorption, whereas CIS results gave uniformly higher transition energies by approximately 2,000cm$^{-1}$, which corresponds to the energy difference between the Fe(II) quasi-ground $a^4F_{9/2}$ state and the free ion $a^6D_{9/2}$ ground state. In light of this experience we use the RPA transition frequencies and strengths to evaluate equation (2). The wavelength of our "UV bump" is therefore from a calculation that has uncertainty of a few percent, based on the prior 480nm comparison of theory and experiment.

The amplitude of the bump depends on the damping constant that we use, in the sense that a smaller damping frequency leads to a narrower optical transition, which has an increased anomalous refractive index that increases extinction (by scattering) to the long wavelength side of a transition. The curves in Figure 9 illustrate the nature of these changes (red curve strongly damped with $\alpha_D$ = 0.2, black curve two times less damped with $\alpha_D$ = 0.1).

The major visible hemoglycin transitions, which depend upon C-termini next to Fe atoms, are included here, but there exists a series of much weaker transitions [McGeoch et al. 2022] associated with other hydroxylation configurations that extends toward the near infrared.

**Discussion**

In an X-ray analysis of the hemoglycin lattice within stromatolite ooids (McGeoch et al 2024a) the above diamond 2H symmetry rod lattice was confirmed, with a slight deviation from exact tetrahedral angles apparently caused by the growth of aragonite crystals within the lattice. In (McGeoch et al. 2021b) we observed that this lattice symmetry had excellent space-filling efficiency defined as the volume of lattice created per rod of the molecule incorporated. This capability was thought to give an advantage to the hemoglycin molecular type in the sense that a larger volume of lattice moving through a molecular cloud would intercept more of the cloud's molecular content that was related to hemoglycin formation, of glycine, iron and silicon, this in turn leading via polymer synthesis to continuing volume expansion. Implicit in this mechanism was a form of replication, or self-assembly, clues for which have been seen in relation to the chiral 480nm absorption of hemoglycin (McGeoch et al 2022). However, the mechanism of this replication remains unknown.

An intuitive measure of space-filling efficiency involved the construction of fictitious volumes related geometrically to the real hemoglycin tetrahedral vertices. It was found that the truncated tetrahedron of Archimedes, a regular solid, when constructed around each hemoglycin lattice vertex, filled 95.8333% of space. We cannot say that this proves the original proposition that out of all structures the rod lattice of diamond 2H symmetry represents the structure that encloses most of three-dimensional Euclidean space per rod of material – our finding is more of a demonstration as to <u>how</u> this very efficient result is achieved by a lattice that exists in nature.

The calculated extinction of hemoglycin dust balls has main absorption features that match astronomical data quite well (Figure 10). Each of these is only present when iron is bonded to a glycine residue, and the remarkable constancy of the bump wavelength is well-explained by the fixed nature of the associated Fe(II) transition. The width of these absorptions cannot be predicted by the present theory, but the prior 51nm (fwhm) measured width of the 483±3 nm feature is consistent with damping coefficients of one tenth the transition frequency, which have been applied here on a trial basis.

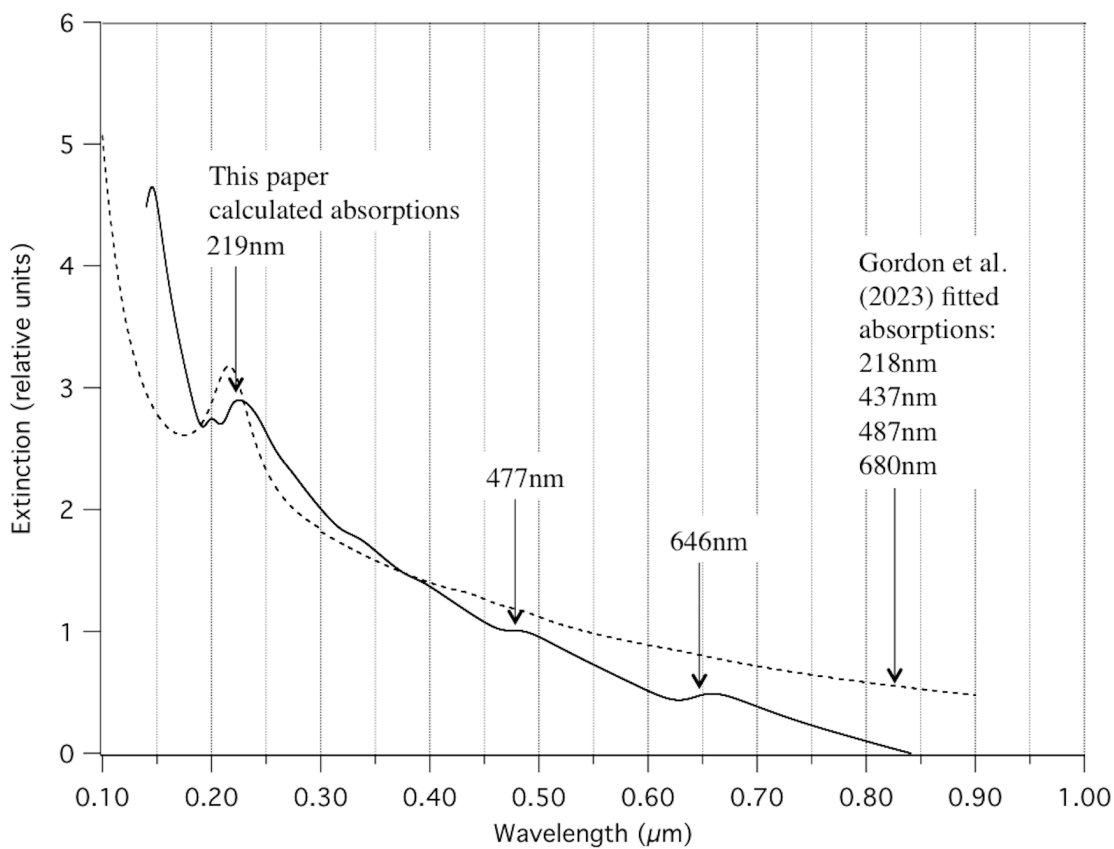

**Figure 10. Calculated hemoglycin lattice extinction (solid curve) compared on a relative basis to a generic extinction fit to astronomical data [Gordon et al. 2023] (dashed curve). At nominal 220nm the peak positions agree within the calculation accuracy.**

The solid curve in Figure 10 has been moved vertically to coincide with the observational curve near the UV bump – only the relative level within each curve is important. Firstly, the overall behavior is very similar over several extinction units which each correspond to a factor of 10. Secondly the calculated 219nm UV bump for hemoglycin (Table 4) is situated at the observed bump within the error of calculation. Thirdly, two subsidiary absorptions that are observationally significant, in the region of 480nm and 650nm, appear on both curves. In hemoglycin the first of these, measured at 483±3nm [McGeoch et al. 2022], originates at an iron atom bonded to the C-terminus of an R-symmetry hydroxyglycine residue, only. The calculated 646nm absorption arises when an iron atom is bonded to the C-terminus of an S-symmetry hydroxyglycine. In summary, the UV bump and the two strongest calculated hemoglycin absorptions in the visible (Table 4) are all present in the generic astronomical extinction curve of [Gordon et al. 2023]. Matching infrared absorptions are discussed elsewhere [McGeoch et al. 2024c].

The calculated far ultraviolet (FUV) extinction below 200nm appears to rise at longer wavelength than astronomical FUV extinction [Gordon et al. 2023]. Below 200nm, convergence difficulties above 40 programmed states due to the increasing density of higher states in Fe forced us to calculate using the simpler no-Fe case of a 22-Glycine antiparallel loop closed at each end by a peptide bond from one strand to the other. This produced absorptions from 146nm to 144nm with a (large) summed transition strength of 5 units, explaining the rapid extinction increase below the bump (numbers in italics in Table 4). However, this part of the calculation is not considered as reliable as that above 200nm because FUV transitions to "conduction" states in the polymer backbone were not included.

In Section S3 it is shown with regard to induced dipoles aligned with lattice rods that the hemoglycin lattice scatters light in exactly the same way as a uniform dielectric medium. This surprising result is attributed to the simplicity and symmetry of the lattice and appears to be related to cosine sum theorems on the uniform irradiation of a sphere by a symmetric set of light beams [Schmitt 1984].

As "dust" the empty hemoglycin lattice with rod length 4.9nm (McGeoch et al 2024a) has a density of 0.030 g/cm$^3$. With increasing densification in a protoplanetary disc it is expected that fractal aggregates of hemoglycin lattice will grow, as discussed for dust in general (Moro-Martin 2019) in relation to the creation of extremely low density fractal aggregates. With hemoglycin the starting density is 33 times smaller than that assumed for icy particles, and the problematic area of "sticking" is taken care of via a "fuzzy" lattice exterior (Figure 5) capable of both ionic and hydrogen bonding, and not necessarily having to exist beyond the "snow-line".

Our one-step extraction technique via cold solvent and MALDI mass spectrometry has allowed us to characterize polymers of amino acids where most studies have boiled meteoritic extracts, derivatized the resultant individual amino acids, and identified them by running against pre-existing samples. We have found that a single core polymer glycine type dominates in these solar system samples, leading us to propose that it could be widespread in the universe, going back to the earliest period that its constituent elements were available following nucleosynthesis (McGeoch et al. 2014). Calculations of absorption and scattering by small lattice balls predict an extinction curve that is a fair representation of the generic observed curve, particularly in relation to the nominal 218nm UV absorption bump. From the ultraviolet to the infrared, the hemoglycin lattice meets the criteria for a universal dust component, not necessarily the only one, but possibly, on occasion, a dominant one.

**Data availability**
All data referred to in the manuscript is published and available via Harvard Dataverse at dataverse.harvard.edu and via request to Julie.mcgeoch@cfa.harvard.edu.

**References**

Aikawa Y., Herbst E., 1999, A&A, 351, 233

Anders E., Grevesse N., 1989, Geochim. et Cosmochim. Acta 53, 197

Blasberger A., Behar E., Perets H. B., Brosch N., Tielens A. G. G. M., 2017, ApJ 836,173

Burton A. S.., Stern J. C., Elsila J. E., Glavin D. P., Dworkin, J. P., 2012 Chem. Soc. Rev., 41, 5459. DOI: 10.1039/c2cs35109a

Dartois E., Noble J. A., Caselli P., Fraser H. J., Jimenez-Serra I., Mate B. et al., 2024, Nature Astronomy, 8, 359

Folch J., Lees M., Sloane Stanley G. H., 1957, J. Biological Chemistry, 226, 497

Gordon K. D., Clayton G. C., Decleir M., Fitzpatrick E. L., Massa D., Misselt K. A. and Tollerud E. J., 2023, ApJ 950, 86  https://doi.org/10.3847/1538-4357/accb59

Iwanami S., Oda N., Radiation Res. 1983, 95, 24  htpss://www.jstor.org/stable/3576068

Jackson J. D., Classical Electrodynamics, Third Edition, 2001, John Wiley and Sons Inc., NJ, USA.

Jiao Y. and Torquato S., 2011, J. Chem. Phys., 135, 151101.

Jones A. P., 1988, MNRAS 234, 209

Jones A. P., Kohler M., Ysard N., Bocchio M. and Verstraete L., 2017, Astronomy and Astrophysics 602, A46

Koga T., Naraoka, H., Scientific Reports, 7, 636 *www.doi.org/*10.1038/s41598-017-00693-9

Mathis J. S., Rumpl W. and Nordsieck K. H., 1977, ApJ 217, 425

McGeoch J. E. M., McGeoch M. W., 2014, PLoS ONE 9(7), e103036. https://journals.plos.org/plosone/article?id=10.1371/journal.pone.0103036

McGeoch J. E. M., McGeoch M. W., 2015, Meteoritics & Planetary Science 50, Nr12 ,1971 https://onlinelibrary.wiley.com/doi/10.1111/maps.12558

McGeoch J. E. M., McGeoch M. W., 2017, https://arxiv.org/pdf/1707.09080.pdf

McGeoch M. W., Šamoril T., Zapotok D., McGeoch J. E. M., 2018, https://arxiv.org/abs/1811.06578

McGeoch M. W., Dikler S. and McGeoch J.E.M., 2021a, https://arxiv.org/abs/2102.10700

McGeoch J. E. M., McGeoch, M. W., 2021b, Physics of Fluids 33, 6, https://aip.scitation.org/doi/10.1063/5.0054860.

McGeoch, J. E. M., McGeoch M. W., 2022,  Scientific Reports 12(1) DOI: 10.1038/s41598-022-21043-4

McGeoch M. W., Owen R. L., Jaho S. and McGeoch J. E. M., 2023, J. Chem. Phys. 158, 114901 https://doi.org/10.1063/5.0143945

McGeoch J. E. M., Frommelt A. J., Owen R., Cinque G., McClelland A., Lageson D. and McGeoch M. W., 2024a, Int. J. Astrobiology 23, e20, 1 https://doi.org/10.1017/S1473550424000168.

McGeoch J.E.M, McGeoch M.W., 2024b, Roy. Soc. Chem. Advances, 14, 36919 https://doi.org/10.1039/d4ra06881e

McGeoch J. E. M., McGeoch M. W., 2024c, arXiv:2309.14914 MNRAS, 530, 1163  https://doi.org/10.1093/mnras/stae756.

Moro-Martin A. 2019, ApJ Lett. 872, L32

Q-Chem. Epifanovsky E. et al., 2021, J. Chem. Phys. 155, 084801 https://doi.org/10.1063/5.0055522

Raffy C., Furthmuller. J. and Bechstedt, F., 2002, Phys. Rev. B 66, 075201 https://doi.org/10.1103/PhysRevB.66.075201

Rojas J., Duprat J., Engrand C., Dartois E., Delauche L., Godard M., Gounelle M., Carillo-Sanchez J. D., Pokorny P. and Plane J. M. C., 2021, Earth Planet. Sci. Lett. 560, 116794  10.1016/j.epsl.2021.116794

Schmitt A. J. 1984, Appl. Phys. Lett. 44, 399

Shimoyama A. and Ogasawara R., 2002, Origins of Life and Evolution of the Biosphere  32, 165 https://doi.org/10.1023/A:1016015319112

Spartan '24, Wavefunction Inc., Irvine CA

Stecher T. P., 1965, ApJ 142, 1683

Stecher T. P., Donn B., 1965, ApJ 142,1681

Stecher T. P., 1969, ApJ 157, L125

https://en.wikipedia.org/wiki/Extinction_(astronomy)

Witstok J., Shivaei I., Smit R. *et al.* (2023) Nature 621, 267  https://doi.org/10.1038/s41586-023-06413-w

# S Section: Space-filling efficiency and optical properties of hemoglycin

Julie E M McGeoch[1] and Malcolm W McGeoch[2]

[1]High Energy Physics DIV, Smithsonian Astrophysical Observatory, Center for Astrophysics, Harvard & Smithsonian, 60 Garden St, MS 70, Cambridge MA 02138. USA.
[2]PLEX Corporation, 275 Martine Str, Suite 100, Fall River, MA 02723, USA

**The empty, extensive low-density lattice topology of hemoglycin is examined to understand how in space, and possibly as early as 800M years into cosmic time a rod-like polymer of glycine and iron came into dominance. A central question to be answered is whether the hemoglycin rod lattice with diamond 2H symmetry represents the most efficient covering of space by a regular arrangement of identical rods. Starting from the tetrahedral symmetry of every hemoglycin lattice we find that the regular truncated tetrahedron of Archimedes may be expanded until neighboring hexagon faces are coincident, at which point space filling is 23/24 or 95.8333% complete. We describe the unit cells of the diamond 2H rod lattice and its conforming near-complete space-filling structure, which has identical symmetry. Maximum space filling via a minimum of molecular material can allow hemoglycin to drive accretion in molecular clouds, contributing to the composition of dust, and providing a background for its widespread presence in meteoritic samples and in cometary material that falls to Earth. The optical properties of hemoglycin lattice balls are derived from quantum calculations of ultraviolet and visible transition energies and strengths. The hemoglycin extinction curve duplicates the nominal 218nm ultraviolet absorption feature known as the UV bump, together with two visible absorption features present in a generic compilation of astronomical extinction data.**

**S1. Isotope Enrichment in hemoglycin**
Mass spectrometry of hemoglycin molecules and fragments throughout the 500 – 2124m/z range has been used to estimate isotope enrichment. This approach, described in (McGeoch et al., 2021a) utilizes the anomalous array of "isotopologues" around a "mono-isotopic" peak to determine a global (+1) mass enrichment relating to either or both of $^2H/^1H$ and $^{15}N/^{14}N$. A separate study involving focused ion beam ejection of negative ions (McGeoch et al. 2018) revealed an enrichment of $^{15}N$ to 1,015 $^0/_{00}$ associated with the -CCN- backbone of polymer amide, a value similar to that of nitrogen in cometary material. In (McGeoch et al., 2021a) we presented $^2H$ enrichment to 25,700 $^0/_{00}$ after a correction of the global enrichment for $^{15}N$. In later work (McGeoch et al., 2024a,b) we simply presented a global (+1) enrichment and provided a conversion method to back out any future $^{15}N$ determinations, to reveal bare $^2H$ numbers.

**Table S1. $^{15}$N and D Isotope evidence that Hemoglycin is ancient in origin (at least 4-5billion years)**

| DATE PUBLISHED | SOURCE | EXPERIMENT | RESULT |
|---|---|---|---|
| 2017 https://arxiv.org/pdf/1707.09080.pdf | Acfer-086 Allende Hawaii volcano control | MALDI/TOF mass spectrometry | 4,641Da molecule with multiples up to 18,600Da $\delta$(global) >10,000 $^0/_{00}$ |
| 2018 https://arxiv.org/abs/1811.06578 | Allende Acfer-086 | FIB of a polymer amide-CCN-backbone into diatomic fragments. | $^{15}$N in the glycine backbone is extraterrestrial (Acfer 086) $\delta^{15}$N = 1,015 $^0/_{00}$ |
| 2021 https://arxiv.org/abs/2102.10700. [physics.chem-ph] | Acfer-086 Allende Kaba | MALDI/TOF MS Very high signal-to-noise achieved | Hemoglycin of mass 1494Da is of glycine, hydroxy-glycine, Fe and O. Hemoglycin is connected covalently in triplets by silicon to form a 4,641Da triskelion. $\delta^2$H = 25,700 $^0/_{00}$ |
| 2024 https://doi.org/10.1017/S1473550424000168 | Fossil and present-day stromatolites | 1. MALDI/TOF MS 2. X-ray diffraction APS and Diamond Light Source synchrotrons 3. FTIR amide band. | Stromatolites contain the same molecule, hemoglycin, as carbonaceous meteorites. $\delta$(global) = 52,000 $^0/_{00}$ **Diamond 2H lattice confirmed in X-ray diffraction.** |
| 2024 https://doi.org/10.1039/d4ra06881e | Sea Foam | MALDI/TOF mass spectrometry | Sea Foam contains the same molecule, hemoglycin, as meteorites and stromatolites $\delta$(global) = 18,300 $^0/_{00}$ |

## S2. Spectroscopic signature of hemoglycin.

**Table S2. Hemoglycin photon absorption/emittance evidence related to telescopic spectra.**

| DATE PUBLISHED | SOURCE | EXPERIMENT | RESULT |
|---|---|---|---|
| 2022<br>10.1038/s41598-022-21043-4 | Acfer-086<br>Sutter's Mill | Measurement of Chiral 480nm absorption | Chiral absorption - strong 483 ± 3 nm |
| 2023<br>https://doi.org/10.1063/5.0143945 | Orgueil | X-ray irradiation of a hemoglycin crystal to produce visible light re-emission. | Visible fluorescence emission to x-rays via Fe-glycine interaction |
| 2024<br>https://doi.org/10.1017/S1473550424000168 | Stromatolites | FTIR and X-ray visible emission | Amide I & Amide II absorption in the region of 6μm |
| 2024<br>https://doi.org/10.1093/mnras/stae756. | Telescopic spectra | Detection by telescope of absorption and emission from 0.2-15μm | Telescope spectra show absorption and emission consistent with hemoglycin |

## S3. Light scattering by the hemoglycin lattice

Here we show that a ball of hemoglycin lattice that is much smaller than the light wavelength scatters light as though from a uniform dielectric sphere – a result that is not intuitive given that in response to an electric field there are different directions of induced polarization in the linear hemoglycin molecules bonded to each other at tetrahedral angles.

We begin by forming a unit cell that contains all of the rod directions in an appropriate distribution. Consider a layer of puckered hexagons as described in Figure 6b of the main text, its six rods each shared by one neighboring hexagon. As discussed, the repeating unit of the lattice is two such hexagons with an upper one rotated $60^0$ relative to a lower one. Combined, within the two hexagons of a unit cell, there are 12 shared rods contributing only 6 unique rod directions. As to the vertical or z-direction connecting rods, between an upper and lower hexagon there are three such connections, each shared by three hexagons intersecting at a vertex, hence only a single connecting rod should be counted for a cell defined by the area of a hexagon. Connection to neighboring cells above and below contributes another one vertical rod (imagined to be half in each of these directions). In summary, there are eight rods per unit lattice cell of volume $32\sqrt{3}h^3/9$ where $h$ is the length of one rod.

We will need their direction cosines to calculate a) the polarizations induced by an applied electric field vector and b) the radiated electric field distribution from the oscillating polarization induced in any one rod. A list of the rod direction cosines is given in Table S3.1 and their labeling is shown in Figure S3.1. In that figure the rods on the hexagon sides appear twice, but as they are shared with the neighboring hexagon, the two are counted as a single rod, giving one rod for each designation $j = 1..8$.

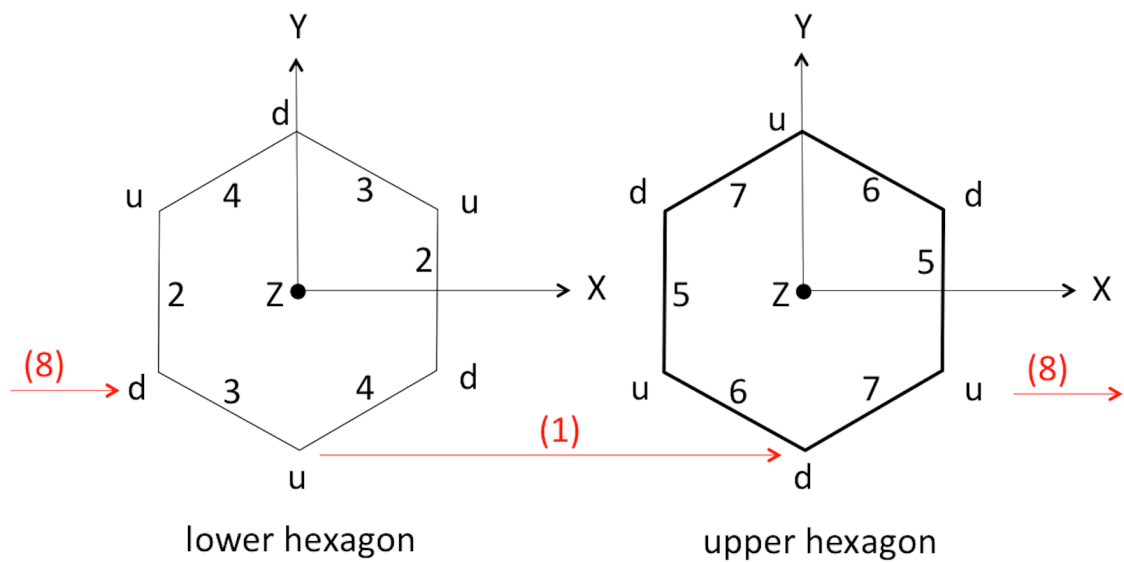

**Figure S3.1 Illustration of the eight rods in a unit cell, where the same orientation has a single index. Because of sharing with neighboring cells, there is only one rod of a given index per unit cell. The verticals in red are all parallel to the z-axis, connecting high vertices (u) in lower hexagons to low vertices (d) in higher hexagons.**

**Table S3.1. Direction cosines for rods numbered as in Figure S3.1, comprising the complete set for a unit lattice cell in diamond 2H symmetry. The vertices belonging to a hexagon are arbitrarily given *u*=up and *d*=down designations in Figure S3.1.**

| Rod designation | x-cosine | y-cosine | z-cosine |
|---|---|---|---|
| 1 | 0 | 0 | 1 |
| 2 | 0 | $\frac{2\sqrt{2}}{3}$ | $\frac{1}{3}$ |
| 3 | $-\sqrt{\frac{2}{3}}$ | $\frac{\sqrt{2}}{3}$ | $-\frac{1}{3}$ |
| 4 | $-\sqrt{\frac{2}{3}}$ | $-\frac{\sqrt{2}}{3}$ | $\frac{1}{3}$ |
| 5 | 0 | $-\frac{2\sqrt{2}}{3}$ | $-\frac{1}{3}$ |
| 6 | $\sqrt{\frac{2}{3}}$ | $-\frac{\sqrt{2}}{3}$ | $\frac{1}{3}$ |
| 7 | $\sqrt{\frac{2}{3}}$ | $\frac{\sqrt{2}}{3}$ | $-\frac{1}{3}$ |
| 8 | 0 | 0 | 1 |

In space the incoming light is un-polarized or partially polarized. Here the discussion may be simplified by considering incident light of a single polarization. Let this light have frequency ω, electric field amplitude $E_0$ and electric field vector direction cosines ($\lambda_E$ $\mu_E$ $\nu_E$). The light is incident on a lattice that is oriented as in Figure S3.1, with trigonal axis parallel to the z-axis. We will calculate the intensity of light scattered in the direction defined by unit vector **n** with direction cosines ($\lambda_N$ $\mu_N$ $\nu_N$). The direction cosines of rod *j* are designated by ($\lambda_j$ $\mu_j$ $\nu_j$). The electric field induced polarization in rod *j* is

$$\vec{P}_j(\lambda_j \mu_j \nu_j) = K \vec{E}_0 (\lambda_E \mu_E \nu_E) \cos\Theta_{EJ}$$

where K is a constant and $\cos\Theta_{Ej} = \lambda_E \lambda_j + \mu_E \mu_j + \nu_E \nu_J$

The scattered (re-radiated) electric field in direction **n** at distance *r* from rod *j* is then [Jackson 2001, equ. 10.2]:

$$\vec{E}_{SCj} = \frac{1}{4\pi\varepsilon_0} k^2 \frac{e^{ikr}}{r} \left[ \left(\vec{n} \times \vec{P}_j\right) \times \vec{n} \right]$$

where $k=2\pi/\lambda$ is the wave vector at frequency ω, and the vector cross product in square brackets specifies the direction of the scattered electric field vector. This holds for a single rod. When the radius *a* of a ball of lattice is much less than the wavelength λ the scattered electric fields can be added in phase to give total field amplitude:

$$\overrightarrow{E_{SC}} = \sum_j \overrightarrow{E}_{SCj} = \frac{1}{4\pi\varepsilon_0} \frac{k^2}{r} \left[ \sum_j (\vec{n} \times \vec{P}_j) \times \vec{n} \right]$$

Also, when $a \ll \lambda$, $N$ unit cells, each of 8 rods, may be added in phase to give, after vector manipulation:

$$\overrightarrow{E_{SC}} = \sum_j \overrightarrow{E}_{SCj} = \frac{N}{4\pi\varepsilon_0} \frac{k^2}{r} \left[ \sum_j \vec{P}_j - \sum_j (\vec{P}_j \cdot \vec{n})\vec{n} \right]$$

The direction and intensity of the scattered radiation energy flow along direction $r$ is

$$\vec{S}(r) = \frac{1}{2} Z_0 |E_{SC}|^2 \vec{n}$$

where $Z_0$ is the impedance of free space.

We note that the scattered intensity depends upon $k^4$ and also upon $N^2$ i.e. the square of the number of unit cells in the lattice ball.

**Results**

The scattered intensity was evaluated to within a constant determined by the factor K above, for various incident inclinations of the electric field to the lattice symmetry axes, and various observation directions. Because we considered a small ball of lattice with $a \ll \lambda$, all the induced polarizations were simultaneously pointed into the half space defined by the direction of the incoming electric field vector. If the arbitrary definitions for the rod directions in Table S3.1 happened give rise to a negative $\cos\Theta_{Ej}$ value, then the direction cosines of that induced polarization were reversed to satisfy this condition. A simple routine was written to evaluate the scattered intensity in terms of a) the propagation direction of an incoming electromagnetic wave relative to the diamond 2H lattice vector set, and b) various polarization directions for the electric field perpendicular to this incident propagation direction. Remarkably, the intensity of scattered light was independent of both (a) and (b), obeying a constant $\sin^2\Theta$ dependence where $\Theta$ is the angle between the incoming propagation direction and the direction of observation of the scattered light. This well known $\sin^2\Theta$ dependence holds for light scattering by a uniform spherical volume of dielectric material when the radius of that volume is much less than the wavelength of the scattered light, as seen in [Jackson 2001, equ. 9.23]:

$$\frac{dP}{d\Omega} = \frac{c^2 Z_0}{32\pi^2} k^4 |\vec{p}|^2 \sin^2\Theta$$

where $dP/d\Omega$ is the power radiated per unit solid angle by an oscillating dipole of moment $\vec{p}$. For a uniform isotropic material of relative dielectric constant $\varepsilon_r$ in a sphere of radius $a$ the total scattering cross section $\sigma$ goes as the sixth power of $a$, representing the coherent superposition of all the internal molecular induced dipoles (for $a \ll \lambda$), viz:

$$\sigma = \frac{8\pi}{3} k^4 a^6 \left| \frac{\varepsilon_r - 1}{\varepsilon_r + 2} \right|^2.$$

The present exercise has shown that a diamond 2H lattice scatters light identically to a uniform isotropic sphere of material.

26